\documentclass[a4paper]{jpconf}

\usepackage[utf8]{inputenc}
\usepackage{amsmath, amssymb, amsfonts}
\usepackage{mathtools}
\usepackage{color}
\usepackage{bm}
\usepackage{graphicx,cancel}
\usepackage{derivative}
\usepackage{calc}
\usepackage{enumitem}
\usepackage{mathrsfs}
\usepackage{multirow}
\usepackage{xspace}
\usepackage{spverbatim}
\usepackage{fancyvrb}
\usepackage{xcolor}
\usepackage{listings}
\usepackage{subfigure}
\usepackage{comment}

\newcommand{\maddm}{{\sc MadDM}\xspace} 
\newcommand{\mg}{{\sc MG5\_aMC}\xspace} 
\newcommand{\madgraphamc}{{\sc MadGraph5\_aMC@NLO}\xspace} 
\newcommand{\fr}{{\sc FeynRules}\xspace}
\newcommand{\fa}{{\sc FeynArts}\xspace}

\newcommand{\nloct}{{\sc NLOCT}\xspace}
\newcommand{\ml}{{\sc MadLoop}\xspace}

\newcommand{\hn}{h}
\newcommand{\Hn}{{H^0}}
\newcommand{\An}{A^0}
\newcommand{\Hp}{{H^\pm}}

\renewcommand{\vec}[1]{\boldsymbol{#1}}

\newcommand{\secvsp}{\vspace{0.8ex}}

\lstdefinestyle{python}{
    language=Python,
    numbers=left,
    numberstyle=\scriptsize\ttfamily,
    stepnumber=1,
    basicstyle=\small\ttfamily,
    keywordstyle=\small\ttfamily\color{blueline},
    stringstyle=\small\ttfamily\color{greenpatch},
    commentstyle=\small\ttfamily\color{gray},
    emph={self,__init__},
    emphstyle=\small\ttfamily\color{redbrickline},
    escapechar=`,
    frame=l
}

\def\slashb#1{\setbox0=\hbox{$#1$}#1\hskip-\wd0\dimen0=5pt\advance
        \dimen0 by-\ht0\advance\dimen0 by\dp0\lower0.5\dimen0\hbox
          to\wd0{\hss\sl/\/\hss}}
\definecolor{ourcyan}{RGB}{0,255,255}
\definecolor{ourmagenta}{RGB}{255,0,255}

\begin{document}
\title{Studying dark matter with \maddm: lines and loops}

%\vspace{-0.5em}

\author{D.~Massaro$^{1,\,2,\,3}$, C.~Arina$^{4}$, J.~Heisig$^{5,\,3}$, F.~Maltoni$^{3,\,1,\,2}$ and O.~Mattelaer$^{3}$}
\address{$^1$~Dipartimento di Fisica e Astronomia, Alma Mater Studiorum - Universit\`a di Bologna, via Irnerio 46, 40126 Bologna, Italy}
\address{$^2$~INFN, Sezione di Bologna, viale Berti Pichat 6/2, 40127 Bologna, Italy}
\address{$^3$~Centre for Cosmology, Particle Physics and Phenomenology (CP3),  Universit\'e catholique de Louvain, Chemin du Cyclotron 2, B-1348 Louvain-la-Neuve, Belgium}
\address{$^4$~Research Institute for Mathematics and Physics (IRMP), Universit\'e catholique de Louvain, Chemin du Cyclotron 2, B-1348 Louvain-la-Neuve, Belgium}
\address{$^5$~Institute for Theoretical Particle Physics and Cosmology, RWTH Aachen University, Sommerfeldstr. 16, D-52056 Aachen, Germany}

\ead{daniele.massaro5@unibo.it}

%\vspace{-0.5em}

\begin{abstract}
Automated tools for the computation of amplitudes and cross sections have become the backbone of phenomenological studies beyond the standard model.
We present the latest developments in \maddm, a calculator of dark-matter observables based on \madgraphamc.
The new version enables the fully automated computation of loop-induced annihilation processes, relevant for indirect detection of dark matter.
Of particular interest is the direct annihilation into photons, $\gamma \gamma,\, \gamma X$.
These processes lead to monochromatic gamma-ray lines that are smoking-gun signatures for dark-matter annihilation in our Galaxy.
\maddm computes the predictions for the expected photon fluxes near Earth and derives constraints from the gamma-ray line searches by Fermi-LAT and HESS\@. As an application, we present the implications for the parameter space of the Inert Doublet Model.
\end{abstract}

\section{Introduction}
\label{sec:intro}
\secvsp

The existence of dark matter has been corroborated by numerous observations over a wide range of physical scales in the Universe.
Indirect detection is one of the main search strategies to shed light on its nature.
It explores traces of dark-matter annihilation (or decay) in cosmic messengers.
In this context, direct annihilation into neutral particles can give rise to prominent spectral features, such as monochromatic lines.
These are smoking-gun signals for dark-matter annihilation, as the pronounced peak is hardly mimicked by any astrophysical background.
In this report, we consider direct annihilation into photons, i.e.~$\gamma \gamma$ and $\gamma X$, where $X = Z, \,h$ or any new particle even under the dark symmetry. These are loop-induced processes given that dark matter is electrically neutral.
The experimental sensitivity to such sharp photon energy spectra is high: current gamma-ray telescopes, like Fermi-LAT~\cite{Ackermann:2015lka} and HESS~\cite{Abdallah:2018qtu}, are able to set strong constraints on the dark-matter parameter space.

Here, we present \maddm v3.2 \cite{Arina:2021gfn}, that features the automatized computation of loop-induced processes for indirect detection within any dark-matter model for which a UFO~\cite{Degrande:2011ua} model at next-to-leading order (NLO) can be generated.
At the time of writing, it is the only numerical tool with such capability.
It further includes the automated computation of the photon fluxes for a variety of dark-matter density profiles and the derivation of exclusion limits by confronting the predictions with the Fermi-LAT satellite and the HESS telescope upper limits.
This new module builds upon the indirect detection module, released with \maddm v3.0~\cite{Ambrogi:2018jqj}.

%While the focus of this new \maddm release is on sharp spectral features, its capability is more general and accounts for {\it any} loop-induced dark-matter annihilation channel: an example is annihilation into  pair of gluons, which lead to a continuum gamma-ray spectrum.
%The new \maddm is designed to recognize the type of energy spectrum produced by the loop-induced process and automatically assign it to the correct analysis pipeline.
%Note that the automatized loop process computation is performed exclusively for indirect detection.

This report is organised as follows.
In \S~\ref{sec:line-pheno} we describe the relevant astrophysical aspects of gamma-ray line searches, in \S~\ref{sec:maddm} we outline the main functionalities of the new \maddm version, while \S~\ref{sec:application} provides a phenomenological study of gamma-ray line signatures within the Inert Doublet Model (IDM).
We conclude in \S~\ref{sec:conclusion}.

\section{Gamma-ray line phenomenology}
\label{sec:line-pheno}
\secvsp

\subsection{Gamma-ray flux and photon energy}
%\subsection{Energy spectrum of the photon line(s)}
\secvsp

The differential flux of gamma rays from annihilation of dark matter in a given region of interest (ROI) is given by\footnote{For non self-conjugate dark-matter candidates eq. \eqref{eq:difflux} has to be multiplied by an additional factor of $1/2$.}
\begin{equation}
\odv{\Phi}{E} = \frac{1}{8 \pi m_\textup{DM}^2} \sum_{i} \langle\sigma v \rangle_i \odv{N^i_\gamma}{E} 
%\int_\textup{ROI} \mathrm{d}\Omega \int_\textup{l.o.s.} \rho^2(\vec{r}) \,\mathrm{d}l \,,
\underbrace{\int_\textup{ROI} \mathrm{d}\Omega \int_\textup{l.o.s.} \rho^2(\vec{r}) \,\mathrm{d}l}_{J\mathrm{-factor}} \,,
\label{eq:difflux}
\end{equation}
where $\langle\sigma v \rangle_i$ is the velocity averaged cross-section of dark-matter particles with a mass $m_\textup{DM}$ into final states labelled by $i$ and 
$\odv{N^i_\gamma}/{E}$ is the respective differential gamma-ray energy spectrum per annihilation.
%The second part of the equation defines the $J$ factor over a region of interest (ROI) in the sky: 
%
%\begin{equation}
%    \label{eq:jfactor}
%    J \equiv  \int_\textup{ROI} \mathrm{d}\Omega\int_\textup{l.o.s.} \rho^2(\vec{r}) \,\mathrm{d}l \,,
%\end{equation}
%
%where $\rho(\vec{r})$ denotes the dark-matter density distribution.
%The second integral is performed over the line of sight (l.o.s.) $l$.
The second part of the equation defines the $J$-factor, where $\rho(\vec{r})$ denotes the dark matter density distribution.
The second integral is performed over the line of sight (l.o.s.)~$l$.
In \maddm, the following dark-matter density profiles are implemented: the Navarro-Frenk-White (NFW) \cite{Navarro:1995iw}, Einasto \cite{Einasto:1965czb}, Burkert \cite{Burkert:1995yz} and isothermal profile \cite{2008gady.book.....B}.

Dark matter is assumed to be non-relativistic in galactic halos: typical values of the dark-matter velocity within the Milky Way \cite{Eilers:2019}, and for similar galaxies, are of the order of $v \simeq 10^{-3} c$.
Accordingly, for dark-matter annihilation into $\gamma\gamma$ or $\gamma X$, the  differential gamma-ray energy spectrum is:
\begin{equation}
    \label{eq:line_spectrum_delta}
    \odv{N_\gamma}{E} = \delta(E - E_{\gamma}) \times 
    \begin{cases}
        2 & \mbox{for } \gamma\gamma \\
        1 & \mbox{for } \gamma X
    \end{cases} \,,
\end{equation}
where $X$ denotes a neutral standard model or beyond standard model (BSM) particle with mass $m_X$ and 
\begin{equation}
    \label{eq:spectrum_line_energy_gammaX}
    E_\gamma = m_\textup{DM} \biggl( 1 - \frac{m_X^2}{4 m_\textup{DM}^2} \biggr) \,.
\end{equation}
This characteristic shape allows for peak searches within the experimental data.

%The most commonly assumed dark matter density profiles are spherically symmetric and given by:
%\begin{itemize}
%    \item generalised Navarro-Frenk-White (NFWg) \cite{Navarro:1995iw} $\rho_{\rm NFW}(r) = \rho_{s} \Bigl(\frac{r_{s}}{r} \Bigr)^{-\gamma} \Bigl(1+\frac{r}{r_{s}} \Bigr)^{\gamma-3}$;
    %%%
%    \item Einasto \cite{Einasto:1965czb} $\rho_{\rm Ein}(r) = \rho_{s} \exp \biggl\{ -\frac{2}{\alpha} \biggl[\Bigl(\frac{r}{r_{s}}\Bigr)^\alpha -1 \biggr] \biggr\}$;
    %%%
%    \item Burkert \cite{Burkert:1995yz} $\rho_{\rm Burkert}(r) = \rho_{s} \Bigl( 1+\frac{r}{r_{s}}\Bigr)^{-1} \biggl[1+ \Bigl( \frac{r}{r_s}\Bigr)^2 \biggr]^{-1}$;
    %%%
%    \item isothermal \cite{2008gady.book.....B} $\rho_{\rm Iso}(r) = \rho_{s} \biggl[1 + \Bigl( \frac{r}{r_s}\Bigr)^2 \biggr]^{-1}$.
%\end{itemize} 
%In all density profiles, the parameters $r_{s}$ and $\rho_{s}$ are the scale radius and the scale density, respectively, normalised to match the specified energy density measured at the Sun position.
%For a given $r_{s}$, $\rho_{s}$ is normalised to match the specified energy density measured at the Sun position (by default $R_\odot = 8.5~\mathrm{kpc}$ and $\rho_\odot = 0.4~\mathrm{GeV}~\mathrm{cm}^{-3}$ is chosen).
%From the NFWg, eq. \eqref{eq:nfwgen}, the usual NFW and the contracted NFW (NFWc) density profiles are obtained for the choice $\gamma=1$ and $\gamma=1.3$, respectively.
%For the Einasto profile, $\alpha$ defines the curvature of the density profile, and it is usually fixed at the value $\alpha=0.17$. 
%
\subsection{Experimental constraints}
\label{ssec:exp_constraints}
\secvsp

Searches for gamma-ray line signals from dark-matter annihilation from the galactic centre have been performed, for instance, by the Fermi-LAT satellite~\cite{Ackermann:2015lka} and the HESS telescope~\cite{Abdallah:2018qtu}.
Fermi-LAT data collected over 5.8 years of observation, provide upper limits in the annihilation cross-section into di-photons constraining dark-matter masses in the range $200~\mathrm{MeV}$ to $500~\mathrm{GeV}$.
%The experimental analysis takes into account four dark-matter density profiles and associates to each one of them an optimised ROI (a circular region with a mask over the galactic plane).
%The ROIs are R3, R16, R41, and R90, which are optimized for the (contracted) NFW, Einasto, NFW, and isothermal density profiles, respectively.
The data collected by the HESS telescope are based on 254 h of live-time observation and are able to constrain rather heavy dark-matter masses, from $300~\mathrm{GeV}$ up to a maximum of $70~\mathrm{TeV}$.
%The analysis is performed for the Einasto density profile and has one optimised ROI, called R1, defined as a circular region centered on the galactic center, with a mask over the galactic plane.

In the case of multiple final states, the application of the experimental constraints requires particular attention.
The photon energy spectrum is a superposition of lines (which in the following will be called {\it peaks}) coming from the various final states $\gamma\gamma$, $\gamma Z$, $\gamma h$, according to eq.~\eqref{eq:spectrum_line_energy_gammaX}.
In the experimental measurement, the predicted gamma-line, a Dirac delta function, is smeared into a Gaussian distribution according to the energy resolution, so that peaks sufficiently close in energy could be indistinguishable for an experiment.
In this case, they should be combined and the final peak's flux would be the sum of the fluxes of each peak.
Furthermore, (combined) spectral lines lacking a sufficient separation from each other question the applicability of the experimental limit-setting procedure, because the experimental analysis has been carried out in the hypothesis of a single photon line.

\section{Loop-induced processes in MadDM}
\label{sec:maddm}
\secvsp

\ml~\cite{Hirschi:2011pa} has been used extensively to compute loop-induced processes in the framework of collider searches for dark matter within \mg~\cite{Hirschi:2015iia}, and can now be used within \maddm~v3.2 to compute loop-induced annihilation processes in galactic halos.
%The task of \ml is to generate all the Feynman diagrams and to numerically evaluate the numerator of such diagram either as a complex number or as a polynomial in the loop-momenta.
%Such information can be used by \textsc{Collier}~\cite{Denner:2016kdg}, \textsc{Ninja}~\cite{Mastrolia:2012bu,Peraro:2014cba,Hirschi:2016mdz}, \textsc{CutTools}~\cite{Ossola:2007ax} or \textsc{IREGI}~\cite{Alwall:2014hca}, to decompose the loop in a sum of scalar integrals either employing tensor integral reduction (TIR, introduced by Passarino \& Veltman \cite{Passarino:1978jh}) or by performing the reduction at the integrand level (OPP method \cite{Ossola:2006us}).
%Finally, the library OneLoop~\cite{vanHameren:2010cp} or QCDLoop \cite{Carrazza:2016gav} are used to return the finite part and the pole of the associated loop.
To be able to perform loop computations within \mg and \maddm it is necessary to import NLO UFO model files. This can be achieved by using \fr~\cite{Alloul:2013bka}, \fa~\cite{Hahn:2000kx} and \nloct~ \cite{Degrande:2014vpa}.

Then, for a given NLO UFO dark-matter model, \maddm automatically generates all contributing diagrams for annihilation into $\gamma X$ with the command:
\begin{verbatim}
    MadDM> generate indirect_spectral_features
\end{verbatim}
where $X$ includes the $Z$ and $h$ particles of the standard model as well as {\it all} additional BSM particles, that are lighter than twice the dark-matter mass and transform even under the dark symmetry that stabilizes the dark matter.
Individual channels can be generated by explicitly specifying the final state, e.g.
\begin{verbatim}
    MadDM> generate indirect_spectral_features a z
\end{verbatim}
The analysis pipeline for these final states starts with the computation of the annihilation cross-section and of the $J$-factor.
Subsequently, it performs a combination (if any) of all peaks to obtain the full gamma-line energy spectrum, finally yielding the prediction of the $\gamma$ flux, and compares it against the current experimental constraints, see~\S~\ref{ssec:exp_constraints}.

Loop-induced annihilation into final states other than the photon ones may contribute to the continuum flux of cosmic messengers.
A relevant example is the dark-matter annihilation into a pair of gluons, $gg$, that subsequently shower and hadronize.
In the absence of a tree-level diagram for this channel, \maddm automatically switches to the loop-induced mode.
Hence, this annihilation channel is considered by executing the command:
\begin{verbatim}
    MadDM> generate indirect_detection g g
\end{verbatim}
It can also be computed together with tree-level diagrams:
\begin{verbatim}
    MadDM> generate indirect_detection
    MadDM> add indirect_detection g g
\end{verbatim}
Here, the first line leads to the computation of all $2\to2$ tree-level annihilation processes.
After the computation of the annihilation cross-section and generation of events, \maddm proceeds with the indirect detection analysis pipeline as introduced in~\cite{Ambrogi:2018jqj}.

\section{Application to the IDM}
\label{sec:application}
\secvsp

To demonstrate the physics impact, in the following we apply the new feature of \maddm to the IDM, that is built on top of the standard model by adding a new (\emph{inert}) Higgs doublet, $\Phi$, odd under an exact $Z_2$ symmetry.
The scalar potential reads
\begin{equation}
	% \begin{multline}
        V = \mu_1^2 \lvert H \rvert^2 + \mu_2^2 \lvert \Phi \rvert^2 + \lambda_1 \lvert H \rvert^4+ \lambda_2 \lvert \Phi \rvert^4
        + \lambda_3 \lvert H \rvert^2 \lvert \Phi \rvert^2 + \lambda_4 \lvert H^\dagger \Phi \rvert^2 + \frac{\lambda_5}{2} \big[ (H^\dagger\Phi)^2 + \mathrm{h.c.} \big] \,.
    % \end{multline}
\end{equation}
%After electroweak symmetry breaking, with $\Phi = (\Hp,\,(\Hn + i \An)/\sqrt{2})^T$, we obtain a total of five physical scalar states with masses given by
%\begin{equation}
%	m_{\hn}^2 = \mu_1^2 + 3 \lambda_1 v^2\,,\quad
%	m_{\Hn}^2= \mu_2^2 + \lambda_L v^2\,, \quad
%	m_{\An}^2 = \mu_2^2 + \lambda_S v^2\,,\quad
%	m_{\Hp}^2 = \mu_2^2 + \frac{1}{2} \lambda_3 v^2\,, 
%\end{equation}
%where $\lambda_\textup{L,S} = ( \lambda_3 + \lambda_4 \pm \lambda_5 ) / 2$.
%
%We are left with five free parameters, after imposing $m_h \simeq 125~\mathrm{GeV}$.
%We express them as \{$m_{\Hn}$, $m_{\An}$, $m_{\Hp}$, $\lambda_L$, $\lambda_2$\}.
After electroweak symmetry breaking, with $\Phi = (\Hp,\,(\Hn + i \An)/\sqrt{2})^T$, we obtain a total of five physical scalar states: $\hn$, $\Hn$, $\An$, $\Hp$. We are left with five free parameters: $m_{\Hn}$, $m_{\An}$, $m_{\Hp}$, $\lambda_L$, $\lambda_2$, where $\lambda_\textup{L,S} = ( \lambda_3 + \lambda_4 \pm \lambda_5 ) / 2$.
Here, we assume $\Hn$ to be the dark-matter candidate and focus on the phenomenologically interesting region around $m_{\Hn} \simeq 72$ GeV.
In this region, the measured relic density can be explained by annihilation into $WW^\ast, ZZ^\ast$ via the gauge kinetic interaction alone. (Here, $V^\ast$ denotes an off-shell vector boson.)
The region is currently unchallenged by current limits from the LHC and direct detection~\cite{Eiteneuer:2017hoh}.

\begin{figure}[t]
     \centering
     \subfigure[Annihilation in $\gamma\gamma$.]{
         \centering
		\includegraphics{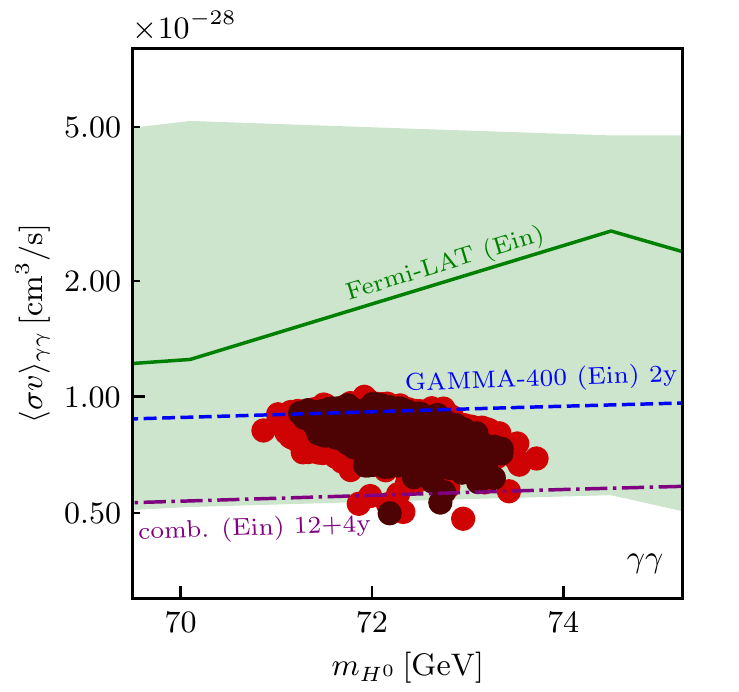}
         \label{fig:IDprospIDM_gammagamma}
     }
     \subfigure[Annihilation in $\gamma Z$.]{
         \centering
         \includegraphics{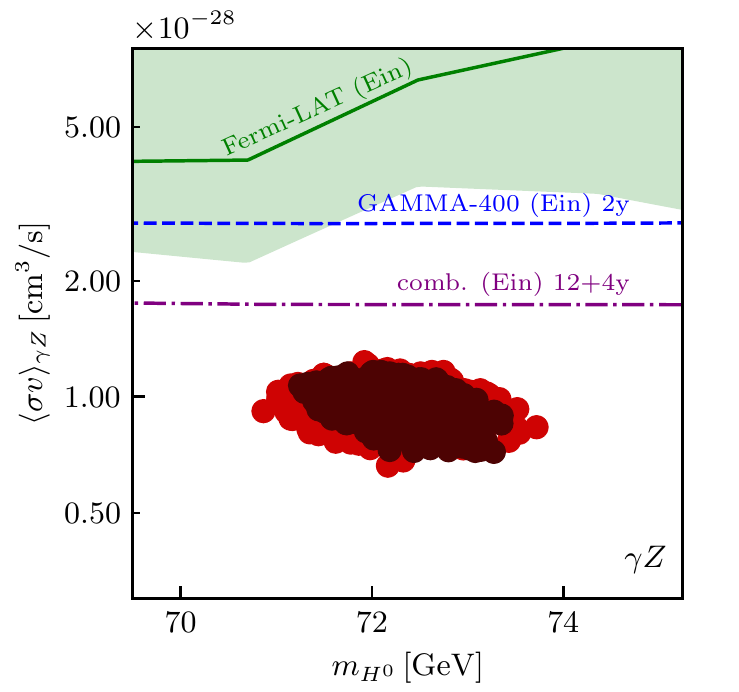}
         \label{fig:IDprospIDM_gammaZ}
        }
        \caption{Cross section for the loop-induced dark-matter annihilation for the allowed IDM parameter points explaining all of dark matter ($\Omega h^2=0.12$) in the range $70\sim75\,$GeV. Parameter points in the $1$ and $2\sigma$ regions are drawn in dark brown and red full circles, respectively. The green solid curve denotes the Fermi-LAT limit for the Einasto profile. The lower and upper boundary of the green shaded band shows the respective limits for NFW contracted and isothermal profile, respectively. The blue dotted and purple dot-dashed curves show projections for GAMMA-400 (2 years) and the combination of Fermi-LAT (12 years) and GAMMA-400 (4 years), respectively, both assuming the Einasto profile.}
        \label{fig:idm_parameter_space}
\end{figure}

We generate the NLO UFO model with \fr, \fa and \nloct and consider the processes $H^0 H^0\to \gamma\gamma$ and $H^0 H^0\to\gamma Z$ involving 140 and 172 diagrams, respectively. 
Note that $H^0 H^0\to \gamma h$ is forbidden due to charge-conjugation invariance.
We have validated our numerical setup using existing results for $\langle\sigma v\rangle_{\gamma\gamma}$ in the literature~\cite{Gustafsson:2007pc,Garcia-Cely:2016hsk} and found agreement within the numerical precision.
We employ the parameter scan performed in~\cite{Eiteneuer:2017hoh} and consider points within the $2\sigma$-region taking into account constraints from the relic density, electroweak precision observables, new physics searches at LEP-II, indirect detection searches for continuous gamma-ray spectra and theoretical requirements of unitarity, perturbativity and vacuum stability. % and performed the gamma-ray line analysis on points within the $2\sigma$ region.
%We considered points within the $2\sigma$ region from the parameter scan and fit performed in~\cite{Eiteneuer:2017hoh}, which takes into account constraints from the relic density~\cite{Ade:2015xua}, electroweak precision observables~\cite{Baak:2014ora,Eriksson:2009ws}, new physics searches at LEP-II~\cite{Pierce:2007ut,Lundstrom:2008ai}, indirect detection searches for continuous gamma-ray spectra from dwarf spheroidal galaxies~\cite{Fermi-LAT:2016uux} and theoretical requirements of unitarity, perturbativity and vacuum stability.
The resulting cross sections for $\langle\sigma v\rangle_{\gamma\gamma}$ and $\langle\sigma v\rangle_{\gamma Z}$ are shown in fig.~\ref{fig:idm_parameter_space} together with the corresponding upper limits from Fermi-LAT~\cite{Ackermann:2015lka} and future projections for GAMMA-400 (2 years) as well as a combination of both observations (with 12 and 4 years observational time, respectively)~\cite{Egorov:2020cmx}.
All lines assume the Einasto dark-matter density profile while the shaded band around the Fermi-LAT limit illustrates the uncertainty due to the choice of the profile.
The upper and lower boundaries of the shaded band corresponds to the upper limit assuming the isothermal and NFWc profiles, respectively.
In the $\gamma\gamma$ channel, the current upper limits only constrain the considered region for the case of NFW (contracted) profile.
For the Einasto profile, future observations are expected to provide sensitivity.
In the $\gamma Z$ channel, the sensitivity of the combination of Fermi-LAT and GAMMA-400 is still too low by about a factor of two to reach the expected signal originating. 

\section{Conclusion}
\label{sec:conclusion}
\secvsp

Gamma-ray lines are smoking-gun signatures of direct dark-matter annihilation into photons in galactic halos.
For electrically neutral dark matter, annihilation into $\gamma\gamma$ and $\gamma X$ is loop-induced.
In this report, we presented \maddm v3.2, that enables the automated computation of loop-induced cross sections for arbitrary dark-matter models implemented in the NLO UFO format via an interface to \ml.
%, optimized for the involved kinematics of non-relativistic dark-matter annihilation.
Furthermore, \maddm computes the resulting integrated photon fluxes and applies experimental constraints from Fermi-LAT and HESS.
It also allows the user to perform the computation of the $J$-factor for a variety of dark-matter density profiles and ROIs.

We demonstrated the capabilities of the program by applying it to the IDM, concentrating on the phenomenologically interesting region of the parameter space $71\,\text{GeV}\lesssim m_\text{DM} \lesssim 74\,\text{GeV}$.
%This region yields the measured relic density without any particular tuning of the involved parameters, providing the required annihilation rate via annihilation into $WW^\ast, ZZ^\ast$ due to the standard-model gauge interactions and it is currently not challenged by direct detection or collider searches.
We found that current limits from gamma-line observation constrain the scenario for the very cuspy NFW (contracted) profile while future observations by GAMMA-400 and Fermi-LAT are expected to probe it for a slightly broader range of dark-matter density profiles.
An intriguing possibility pointing to the IDM model would be the observation of two peaks, from $\gamma\gamma$ and $\gamma Z$, which are well separated for the considered dark-matter masses.
However, the $\gamma Z$ signal still appears out-of-reach according to the above-mentioned projections. 

%This release can be seen as the first step towards incorporating automated loop-level computations in \maddm.
%In the light of current data from the cosmic microwave background enabling a relic density measurement below percent level precision, higher-order corrections in the corresponding theoretical predictions become relevant and call for automated computational tools.
%While \mg utilizes a suitable framework for this quest, an efficient computation of thermally averaged cross sections at NLO requires further research that is left for future work.

\section*{References}
\secvsp

\bibliography{bibliography}
\end{document}